\documentclass[12pt]{article}

\usepackage{graphicx}
\usepackage{graphicx}

\def\bb{\begin{equation}}
\def\ee{\end{equation}}
\def\pt{\partial}

\def\ve{\varepsilon}

\newtheorem{theorem}{Theorem}

\title{{\bf The autoresonance threshold into a system of weakly coupled oscillators}
\thanks{This work was supported by grants RFBR 06--01--00124, 06--01--92052-KE and INTAS 03-51-4286.}}

\author{Sergei Glebov\thanks{Ufa State Petroleum Technical University; 
glebskie@gmail.com},\quad Oleg Kiselev\thanks{Institute of Math. USC
RAS; ok@ufanet.ru},\quad
Vladimir Lazarev\thanks{Ufa State Petroleum Technical University; 
lazva@mail.ru}}

\date{}

\begin{document}

\maketitle

\begin{abstract}
We investigate a system of two weakly coupled oscillators. It is shown that an external periodic perturbation can lead to the capture into resonance. Asymptotic description and numerical simulations are presented. We have obtained the explicit formula for the threshold value for the amplitude of the perturbation that leads to the autoresonance.
\end{abstract}

\section*{Introduction}

A system of nonlinear oscillators  is often used as a standard model for oscillating  processes. The oscillators have  eigen-frequencies and can resonantly  interact between each other or with an external perturbation. 
A slow variation of frequencies can lead to the increase of amplitudes of oscillations.
 This phenomenon is usually called the capture into autoresonance and was found in \cite{Mc}, \cite{Veksler1}, \cite{Veksler2}. Later similar effects were observed in various  fields of modern physics  \cite{MeerFr}, \cite{MeerYr},\cite{MeerCo},\cite{Fr}.
\par
Various aspects of autoresonance solutions were investigated for one-dimensional problems \cite{KG1}, \cite{KG2} and for multi-dimensional problems \cite{KB}, \cite{LK3}. 
\par
In this article we investigate the system of two weakly coupled nonlinear oscillators under a small perturbation 
\begin{eqnarray}
x''+ \omega^2 x &=&\ve \alpha_1 xy + \ve(\gamma\exp\{i\varphi\} + c.c.), \label{sys1}\\
y''+ (2\omega)^2 x &=&\ve \alpha_2 x^2, \nonumber
\end{eqnarray}
here $\varphi = (\omega +\ve \alpha\tau)\theta, \ \tau =\ve \theta$, $\ve$ is a small positive parameter.
\par
The constants in the system have the following sense $\omega=const$ is frequency of the oscillator with the amplitude $x$, $\alpha_1$ and $\alpha_2$ are parameters of the nonlinear coupling,  $\gamma$ is the amplitude of the external perturbation, $\alpha$ is the derivative of detuning of frequency of external perturbation  with respect to slow time $\tau$.
\par
When $\ve=0$ the eigen-frequencies  relate to each other as $1:2$. If $\ve\not=0$ then this relation leads to the parametric resonance. When external perturbation $\gamma=0$ the nonlinear oscillators exchange energy \cite{Abl},\cite{Ruben} and the order of solutions does not change. 
\par
Our goal is to obtain a solution with an increasing amplitude due to a small oscillating perturbation when $\gamma\not=0$. There is a standard way to obtain oscillations  with an increasing amplitude. You should use  the slow varying frequency of the perturbation and the effect of autoresonance.
\par
Not long ago it was found that this simple receipt is not sufficient for the capture into autoresonance. It was numerically obtained that there is a threshold value of the perturbation when the capture into resonance is not observed  for subharmonic modes \cite{Fr1}, \cite{Fr2}. Later the similar results were obtained for  an oscillator with a quadratic form with respect to $\tau$ for the phase of the perturbation \cite{LK1}, \cite{LK2}. 
\par
In this article we have obtain a new result for two-dimen\-sional system of  primary resonance equations.
 It was shown that the autoresonant phenomenon appears when the amplitude of the perturbations is greater than the threshold value. This threshold value of the perturbation was found explicitly.
\par
The mathematical statement of the problem on the existence of threshold is reduced to an existence of asymptotic solutions in the form of power series with respect to $t^{-1}$ as $t\to \infty$. The general approach to the construction of power asymptotics can be read in  \cite{FK}, \cite{BRUNO}. However the direct using of 
those methods does not allow one to predict the existence of  threshold phenomenon in autoresonance equations.
\par

\section{Statement of the problem and result}

The long time evolution of amplitudes of oscillating solutions of   (\ref{sys1}) is reduced to  the system of primary resonance equations:
\begin{eqnarray}
A'(t) &=& -i\left(2 t A + \frac{1}{2}A^*B +f \right), \nonumber \\
B'(t) &=& -i\left( 4tB + \frac{1}{4}A^2\right). \label{sys3}
\end{eqnarray}
\par
Our goal is to study the behaviour of the solutions of (\ref{sys1}) when $t$ approaches infinity.
\par
It is shown there are increasing and bounded solutions. The bounded solutions are studied in detail. It has been determined that periodic perturbation of a system of  oscillators leads to the capture into resonance. The asymptotic description and numerical simulations of the phenomenon are presented. An explicit formula for the threshold value of the perturbation has been found. It was found  there exist solutions related to autoresonance phenomenon when $|f|\ge 12$.
\par

\section{Asymptotic reduction to the system of  primary resonance equations}

In this section we present a formal asymptotic reduction of a system of  primary resonance equations.
Let us construct the solution for (\ref{sys1}) in the complex form
\bb
x={\cal A}(\tau)\exp\{i\omega \theta\} + c.c., \qquad y={\cal B}(\tau)\exp\{2i\omega \theta\} + c.c. \label{anzats1}
\ee
Substitute (\ref{anzats1}) into (\ref{sys1}). It yields
\begin{eqnarray}
{\cal A}'(\tau) &=&  - \frac{i\alpha_1}{2\omega}{\cal A}^*{\cal B} - \frac{i\gamma}{2\omega}\exp\{i\alpha\tau^2\}+i\ve{\cal A}'' , \nonumber \\
{\cal B}'(\tau) &=&  -\frac{i\alpha_2}{4\omega}{\cal A}^2 +i\ve{\cal B}''.  \nonumber
\end{eqnarray}
Neglecting the terms of order $\ve$ and substituting
\bb
{\cal A}=a(\tau)\exp\{i\alpha \tau^2\}+c.c.,\quad {\cal B}=b(\tau)\exp\{2i\alpha \tau^2\}+c.c. \label{anzats11}
\ee
we obtain
\begin{eqnarray}
a'(\tau) &=& -2i\alpha \tau a - \frac{i\alpha_1}{2\omega}a^*b - \frac{i\gamma}{2\omega}, \nonumber \\
b'(\tau) &=& -4i\alpha \tau b -\frac{i\alpha_2}{4\omega}a^2. \label{sys2}
\end{eqnarray}
\par
Let us simplify system (\ref{sys2}).
Substitution
$$
a(\tau) = \lambda A(t),\quad b(\tau) = \kappa B(t),\quad \tau = \chi t,
$$
where
$$
\kappa = \frac{\omega\sqrt{\alpha}}{\alpha_1}, \quad \lambda = \omega\sqrt{\frac{\alpha}{\alpha_1\alpha_2}}, \quad \chi^2 = \frac{1}{\alpha}
$$
gives  system (\ref{sys3}) with $f=\displaystyle\frac{\sqrt{\alpha_1\alpha_2}\gamma}{\alpha\omega^2}$. 
\par

\section{Numerical simulations}

In this section we present results of numerical simulations for system (\ref{sys3}). Figures 1 and 2 show 
 $|A(t)|$ for the captured and noncaptured solution. Initial data were identical for both cases 
$$
A(100) = 102.669-i 793.88, \quad B(100) = 386.825+i101.831.
$$
We have used $f=11.9$ for the first figure and $f=12.1$ for the second one. The second figure corresponds to the captured solution where the amplitude of the perturbation is greater than the threshold value.
\par
\vskip0.5cm
\begin{tabular}{cc}
\includegraphics[width=5cm,height=4cm]{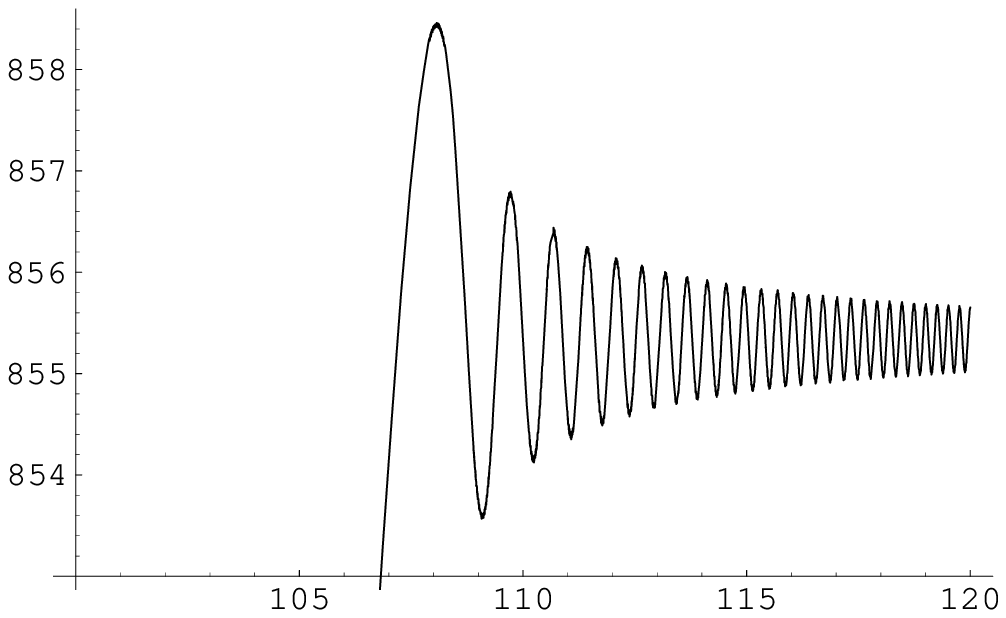}%for DVI-file
& 
\includegraphics[width=5cm,height=4cm]{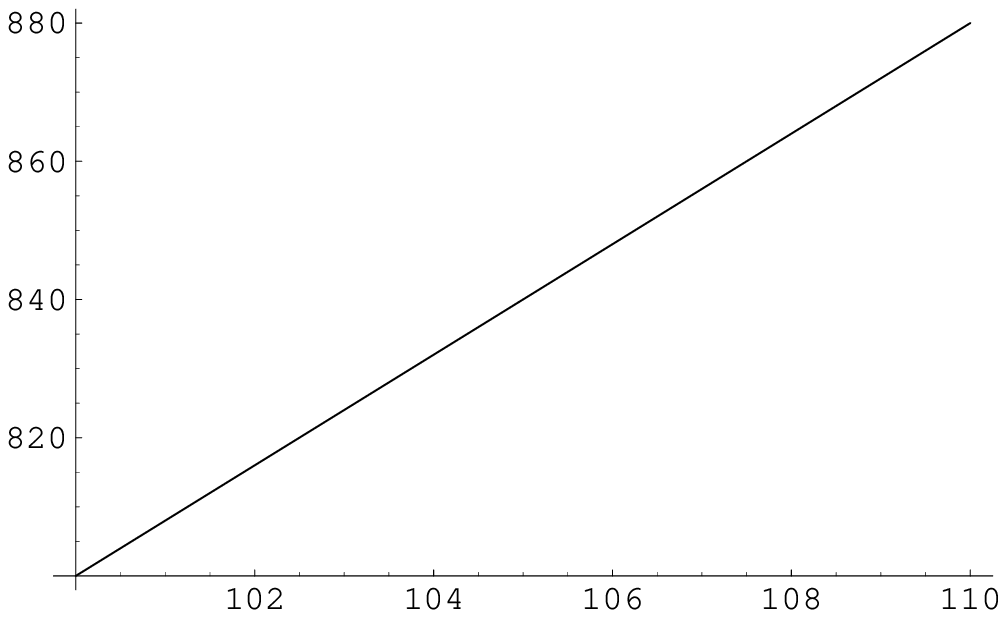} \\ %for DVI-file
Fig.1: noncaptured solution & Fig.2: captured solution
\end{tabular}
\par
\vskip0.5cm
Figures for $|B(t)|$ look similar for both cases. The presented numerical results show that there exists the threshold value of  amplitude of perturbation that leads to essential change of structure of solution.
\par

\section{Algebraic asymptotic solutions}

In this section we construct algebraic asymptotic solutions of system (\ref{sys3}) in the form 
\bb
A(t) = \sum_{k=-1}^\infty a_{k} t^{-k}, \quad 
B(t) = \sum_{k=-1}^\infty b_{k} t^{-k}, \quad t\to\infty. \label{algAsymp} 
\ee
\begin{theorem}
When $t\to\infty$ there exists the solution of system  (\ref{sys3}) of the form
\begin{eqnarray*}
A_2(t) &=& -\frac{f}{2}t^{-1} +\frac{if}{4}t^{-3} + \Big(\frac{3f}{8} - \frac{f^3}{512}\Big) t^{-5} + O(t^{-7}),  \\
B_2(t) &=& -\frac{f^2}{64}t^{-3} + \frac{7if^2}{256}t^{-5} + O(t^{-7}), 
\end{eqnarray*}
When $|f|\ge 12$ and $t\to\infty$  there exist solutions of (\ref{sys3}) of the form
\begin{eqnarray*}
A_1(t) &=& -8\Big(\cos(\Psi)+i\sin(\Psi)\Big)t + \frac{f}{4}t^{-1} + O(t^{-3}),  \\
B_1(t) &=& -4\Big(\cos(2\Psi)+i\sin(2\Psi)\Big)t +(-\frac{f}{4}-2i)t^{-1} + O(t^{-3}), 
\end{eqnarray*}
here $\sin(\Psi)=\displaystyle\frac{12}{f}$.
\begin{eqnarray*}
A_3(t) &=& 8\Big(\cos(\Psi)+i\sin(\Psi)\Big)t + \frac{f}{4}t^{-1} + O(t^{-3}),  \\
B_3(t) &=& -4\Big(\cos(2\Psi)+i\sin(2\Psi)\Big)t  + \Big(-\cos(\Psi)[\frac{f}{4}+\frac{24}{f}]+2i[1+\sin^2(\Psi)]\Big)t^{-1}  \\ 
 &+& O(t^{-3}),
\end{eqnarray*}
here $\sin(\Psi)=-\displaystyle\frac{12}{f}$.
\end{theorem}
The proof of the Theorem consists in the construction of the asymptotic solution of (\ref{sys3}) and using Kuznetsov's Theorem \cite{Kuz}.
\par
Let us substitute representation (\ref{algAsymp}) into (\ref{sys3}) and gather the terms of the same order of $t$. It gives the recurrent system of equations for coefficients $a_k, b_k$ of asymptotic expansion (\ref{algAsymp})
\begin{eqnarray}
2ia_{k}+\frac{i}{2}\left(a^*_{k}b_{-1}+ a^*_{-1}b_{k}\right) =(k-2)a_{k-2}-\frac{i}{2}\sum_{m,l} a^*_m b_l,\nonumber \\
4ib_{k}+\frac{i}{2}a_{k}a_{-1} = (k-2)b_{k-2} -\frac{i}{4}\sum_{m,l} a_m a_l,\label{problemsCoef}
\end{eqnarray}
here $m+l=1-k$ and $m,l\not=k$.
\par
Terms under $t^2$ give
$$
2a_{-1}+\frac{1}{2}a^*_{-1}b_{-1} =0, \quad 4b_{-1} +\frac{1}{4}a^2_{-1} =0.
$$
Coefficient $a_{-1}$ is the solution of equation
$$
\frac{1}{32}|a_{-1}|^2a_{-1} -2a_{-1} =0.
$$
It shows there are two increasing solutions with $|a_{-1}|^2 = 64$ and one bounded solution with $a_{-1}=0$. The leading-order term $b_{-1}=-\displaystyle\frac{1}{16}a^2_{-1}$.
\par
Here we give the procedure of construction of increasing solution that goes to $+\infty$ when t also increases. Another increasing solution that goes to $-\infty$ will be discussed below.
\par
We suppose that
$$
a_{-1} = 8 \exp\{i\Psi\}, \quad b_{-1} = - 4 \exp\{2i\Psi\}.
$$
\par
Terms of order of $t$ give a homogeneous system for $a_0$ and $b_0$. The matrix of the system for real and imaginary  parts of $a_0$ and $b_0$ has the form
\bb\left[
\begin{array}{cccc}
4\cos(\Psi)\sin(\Psi) & -4\cos(\Psi)^2 & 4\sin(\Psi) & -4\cos(\Psi) \\
4\sin(\Psi)^2 & -4\cos(\Psi)\sin(\Psi) & 4\cos(\Psi) & 4\sin(\Psi) \\
-4\sin(\Psi) &  -4\cos(\Psi) & 0 & -4 \\
4\cos(\Psi) & -4\sin(\Psi) & 4 & 0
\end{array}\right]\label{matrix0}
\ee
The rang of the matrix equals three and solution depends on a parameter. The solution is
$$
Y=\mu_0Y_0 = \mu_0\Big(\sin(\Psi),-\cos(\Psi),-\sin(2\Psi),\cos(2\Psi)\Big).
$$
It yields 
$$
a_0 = \mu_0\Big[\sin(\Psi) - i\cos(\Psi)\Big],\quad b_0 = \mu_0\Big[-\sin(2\Psi) + i\cos(2\Psi)\Big].
$$
Terms of order of $t^0$ give a nonhomogeneous system for real and imaginary  parts of $a_1$ and $b_1$ with the matrix (\ref{matrix0}). To obtain a non-trivial solution the right-hand side 
\begin{eqnarray*}
F=\Big(-8\cos(\Psi) - \frac{1}{2}\mu_0^2\sin(\Psi),-f + \frac{1}{2}\mu_0^2\cos(\Psi)- 8\sin(\Psi),  \\
 4\cos(\Psi)^2 -\frac{1}{2}\mu_0^2\cos(\Psi)\sin(\Psi) - 4\sin(\Psi)^2, \\
  \frac{1}{4}\mu_0^2\cos(\Psi)^2 + 
8\cos(\Psi)\sin(\Psi) - \frac{1}{4}\mu_0^2\sin(\Psi)^2\Big)
\end{eqnarray*}
should be orthogonal to the solutions of the adjoint  system. The solution of adjoint system is
\bb
Z=\Big(-\cos(\Psi),-\sin(\Psi),\cos(2\Psi),\sin(2\Psi)\Big). \label{solCon}
\ee
The solvability condition for the system for real and imaginary  parts of $a_1$ and $b_1$ has the form
\bb
\sin(\Psi) = -\frac{12}{f}. \label{porog}
\ee
The variable $\Psi$ that determines a turning of the leading-order term of asymptotic expansion can be determined  when 
$|f|\ge 12$.
\par
Real and imaginary parts of $a_1$ and $b_1$ are represented as a sum of general solution for homogeneous equations and partial solution of nonhomogeneous equation.
\begin{eqnarray*}
(\Re[a_1],\Im[a_1],\Re[b_1],\Im[b_1]) = \mu_1Y + \Big(\frac{4\cos(\Psi)-\mu_0^2}{16\cos(\Psi)}, 0, \\
 \frac{\cos(\Psi)(-192-2f^2+f\mu_0^2\cos(\Psi))}{8f}, \\ \frac{1152f\cos(\Psi)+8f^3\cos(\Psi)+864\mu_0^2-9f^2\mu_0^2}{4f^3\cos(\Psi)}\Big)^{T}
\end{eqnarray*}
Coefficients for higher-order terms of asymptotics are determined in the similar way. The solutions for these coefficients depend on a parameter $\mu_k$. This parameter is determined from the solvability condition for the nonhomogeneous system of algebraic equations with degenerate matrix. The solvability condition looks as follows
\bb
Z\cdot (F_1,F_2,F_3,F_4) =0.\label{ortCond}
\ee
where $Z$ is a solution of the adjoint system, $(F_1,F_2,F_3,F_4)$ is the right-hand side of equation for real and imaginary part of $a_{k+2}, b_{k+2}$.
\par
Vector $(F_1,F_2,F_3,F_4)$ contains a product of coefficients for higher-order correction terms of (\ref{algAsymp}). The algebraic system is written for 
$$
\Big(\Re[a_k], \Im[a_k],\Re[b_k], \Im[b_k]\Big).
$$
It yields the following rule for right-hand side 
\begin{eqnarray}
F_1 =-\Re[a_{k-1}]+\Re\Big[\sum a^*_m b_l\Big], \nonumber \\
F_2 =-\Im[a_{k-1}] + \Im\Big[\sum a^*_m b_l\Big], \nonumber \\
F_3 =-\Re[b_{k-1}] + \Re\Big[\sum a_m a_l\Big], \nonumber \\
F_4 =-\Im[b_{k-1}] + \Im\Big[\sum a_m a_l\Big]. 
\end{eqnarray}
The parameter $\mu_k$ is determined from relation (\ref{ortCond}) for all values of $k$. The equation for $\mu_k$ looks as follows
\bb
Z\cdot (Y_0 \odot X_k) - Z\cdot X_{k-1}=0, \label{vProduct}
\ee
where $Z$ is a solution of homogeneous union system, $Y_0$ is a solution of homogeneous system and $X_k$ is a partial solution of nonhomogeneous system for $k$-th correction term. An operation $\odot$ is determined by
$$
Y \odot X = \left[\begin{array}{c} y_3x_1 + y_4x_2 + y_1x_3 + y_2x_4 \\
                            - y_3x_2 + y_4x_1 - y_4x_3 + y_1x_4 \\
                            	 2y_1x_1 -2y_2x_2 \\
                            	  2y_2x_1 + 2y_1x_2  \end{array}\right].
$$
We determine the parameter $\mu_k$ and relation (\ref{vProduct}) becomes valid. 
\par
The solution  $X_k$ can be expanded on a basis $Y_0, Y_1, Y_2, Y_3$, where
\begin{eqnarray}
Y_0&=&\Big(\sin(\Psi),-\cos(\Psi),-\sin(2\Psi),\cos(2\Psi)\Big),\nonumber \\
Y_1&=&\Big(\cos(\Psi),\sin(\Psi),0,0\Big),\nonumber \\
Y_2&=&\Big(0,0,\cos(2\Psi),\sin(2\Psi)\Big),\nonumber \\
Y_3&=&\Big(0,0,-\sin(2\Psi)\cos(2\Psi),\cos(2\Psi)\sin(2\Psi)\Big).
\end{eqnarray}
Direct calculations give
$$
Z\cdot (Y_0 \odot Y_i)=0, \hbox{for}\ i=0,1\quad \hbox{and}\quad Z\cdot (Y_0 \odot Y_i)\not=0, \hbox{for}\ i=2,3,
$$
and the vector $Z$ is not orthogonal to vectors $Y_1, Y_2, Y_3$. It yields a nontrivial equation for  $\mu_k$
$$
\mu_k Z\cdot\Big(Y_0 \odot [C_{k,2}Y_2+C_{k,3}Y_3]\Big)=Z\cdot\Big(C_{k-1,1}Y_1+C_{k-1,2}Y_2+C_{k-1,3}Y_3 \Big).
$$
\par
{\bf Note.} We can construct another solution with the leading-order term $a_1 = -8$ in similar way. The change of  sign for $a_1$ does not lead to essential change of  procedure of asymptotic construction.
\par
Above asymptotic constructions give the following algebraic asymptotic expansions for increasing solutions
\begin{eqnarray}
A_1(t) &=& -8\Big(\cos(\Psi)+i\sin(\Psi)\Big)t + \frac{f}{4}t^{-1} + O(t^{-3}), \nonumber \\
B_1(t) &=& -4\Big(\cos(2\Psi)+i\sin(2\Psi)\Big)t +(-\frac{f}{4}-2i)t^{-1} + O(t^{-3}), \label{asMinusInfty}
\end{eqnarray}
where $\sin(\Psi)=\displaystyle\frac{12}{f}$.

\begin{eqnarray}
A_3(t) &=& 8\Big(\cos(\Psi)+i\sin(\Psi)\Big)t + \frac{f}{4}t^{-1} + O(t^{-3}),  \\
B_3(t) &=& -4\Big(\cos(2\Psi)+i\sin(2\Psi)\Big)t  + \Big(-\cos(\Psi)[\frac{f}{4}+\frac{24}{f}]+2i[1+\sin^2(\Psi)]\Big)t^{-1}  \nonumber  \\ 
&+& O(t^{-3}), \label{asPlusInfty}
\end{eqnarray}
where $\sin(\Psi)=-\displaystyle\frac{12}{f}$.
\par
We construct the finite solution of the form
\begin{eqnarray}
A(t) &=& \sum_{k=1}^\infty a_{k} t^{-k}, \nonumber  \\
B(t) &=& \sum_{k=1}^\infty b_{k} t^{-k}. \label{algAsymp0} 
\end{eqnarray}
Substitution of (\ref{algAsymp0}) into equation gives the recurrent sequence of problems  (\ref{problemsCoef}) for coefficients of (\ref{algAsymp0}). The coefficients $a_{k}$ and $b_{k}$ are determined from a system of algebraic equations with a nondegenerate matrix. 
\par
Relations of order of $t^{0}$ give the following system for real and imaginary parts of the leading-order terms
$$
2ia_{1r}-2a_{1i} = -if,\quad 4ib_{1r} -4b_{1i=0}.
$$
These equations allow us to determine the leading-order term of asymptotic expansion
$$
a_1 =-\frac{f}{2},\quad b_1=0.
$$
The real and imaginary parts of $a_2, b_2$ should equal zero.
\par
Relations of order of $t^{-2}$ give nontrivial equations 
$$
2ia_{3r} -2a_{3i} = -\frac{f}{2},\quad 4ib_{3r} - 4b_{3i} = -\frac{if^2}{16}.
$$
It yields
$$
a_3 =\frac{if}{4},\quad b_3=-\frac{f^2}{64}.
$$
The recurrent procedure gives
\begin{eqnarray}
A_2(t) &=& -\frac{f}{2}t^{-1} +\frac{if}{4}t^{-3} + \Big(\frac{3f}{8} - \frac{f^3}{512}\Big) t^{-5} + O(t^{-7}), \nonumber \\
B_2(t) &=& -\frac{f^2}{64}t^{-3} + \frac{7if^2}{256}t^{-5} + O(t^{-7}), \label{asZero}
\end{eqnarray}

\section{Neighborhoods of equilibrium positions}
\subsection{Stability by  linear approximation}
\par
In this section we present the analysis of stability for the constructed algebraic asymptotic solutions by  linear approximation.
\par 
Let us consider the system of equations that are linearized on the increasing asymptotic solution $(A_1(t),B_1(t))$.
Substitute 
$$
a(t)= A_1(t)+ \alpha(t),\quad b(t)=B_1(t)+\beta(t),
$$
into (\ref{sys3}). It gives the system of linear differential equations for $\alpha(t)$ and $\beta(t)$. It is convenient  to rewrite this system as a system for real and imaginary parts of $\alpha(t)$  and $\beta(t)$. The eigen-values for the system are
$$
\lambda_1=-4 i\sqrt{3}t+O(t^{-1}),\qquad \lambda_2=4 i\sqrt{3}t+O(t^{-1}),
$$
$$
\lambda_3=-\frac{\sqrt[4]{f^2-144}}{\sqrt{6}}+O(t^{-1}),\qquad \lambda_4=\frac{\sqrt[4]{f^2-144}}{\sqrt{6}}+O(t^{-1}).
$$
$\lambda_4$ has a positive real part when $f>12$. It yields that the constructed above algebraic solution is not stable with respect to small perturbations.
\par
The similar linearization on the algebraic solution $A_2(t),B_2(t)$ leads to the matrix with eigen-values of form
$$
\lambda_1=-4 i t+O(t^{-1}),\qquad \lambda_2=4 i t+O(t^{-1}),
$$
$$
\lambda_3=-2it+O(t^{-1}),\quad \lambda_4=2it+O(t^{-1}).
$$
The leading-order terms of the given asymptotic expansions are imaginary and  a question on stability of  $A_2(t),B_2(t)$ requires additional investigations.
\par
A matrix of the system of linear differential equations that are realized on $A_3(t),B_3(t)$ has eigen-values of the form
$$
\lambda_1=-4 i t+O(t^{-1}),\qquad \lambda_2=4 i t+O(t^{-1}),
$$
$$
\lambda_3=-i\frac{\sqrt[4]{f^2-144}}{\sqrt{6}}+O(t^{-1}),\qquad \lambda_4=i\frac{\sqrt[4]{f^2-144}}{\sqrt{6}}+O(t^{-1}).
$$
It leads to additional investigations of stability of $A_3(t),B_3(t)$ for $f>12$.

\subsection{Oscillating asymptotic solution in the neighborhood of bounded solution}

In this section we present an oscillating asymptotic solution in the neighborhood of bounded solution.
The constructed solution of (\ref{sys3}) has the form  
\bb
A(t)=a(t)\exp\{-it^2\}, \quad B(t)=b(t)\exp\{-2it^2\}. \label{WKBanzat}
\ee
The substitution gives
\begin{eqnarray}
ia'(t) &=& \frac{1}{2}a^*b +f\exp\{it^2\} , \nonumber \\
ib'(t) &=& \frac{1}{4}a^2, \label{WKBsys}
\end{eqnarray}
Here we study the behavior of solution of the system in the neighborhood of bounded asymptotic solution  $(A_2,B_2)$. 
Substitution
\bb
a = A_2\exp\{it^2\} + \alpha ,\quad   b = B_2\exp\{2it^2\} + \beta, \label{asWKB0}
\ee
gives a system for $\alpha, \beta$
\begin{eqnarray}
i\alpha'=\frac{1}{2}\alpha^*\beta + \frac{1}{2}(A_2^*\beta\exp\{-it^2\} + \alpha^*B_2\exp\{2it^2\}), \nonumber\\
i\beta'=\frac{1}{4}\alpha^2 + \frac{1}{2}A_2\alpha\exp\{it^2\}. \label{sysWKB1}
\end{eqnarray}
\begin{theorem}
There exists a formal asymptotic solution of (\ref{sys3}) in the form of (\ref{asWKB0}) with 
\bb
\alpha = \sum_{k=0}^{\infty} \alpha_k t^{-k}, \quad \beta = \sum_{k=0}^{\infty}\beta_k t^{-k}. \label{coeffs}
\ee
This asymptotic solution  depends on four real parameters. The leading-order terms of the expansion are determined in terms of elliptic functions.
\end{theorem}
The same result for the system of three coupled oscillators was obtained in \cite{KB}.
\par
The proof of the theorem will be obtained by formal asymptotic construction of (\ref{coeffs}). Substitute (\ref{asWKB0}), (\ref{coeffs}) into (\ref{sysWKB1}) and take into account  (\ref{asZero}). Gathering the terms of the same order of $\tau$ gives a recurrent system of equations for coefficients of the asymptotic expansion. The leading-order terms are satisfied to 
\begin{eqnarray}
i\alpha_0'(t) &=& \frac{1}{2}\alpha_0^*\beta_0  , \nonumber \\
i\beta_0'(t) &=& \frac{1}{4}\alpha_0^2. \label{WKBleading}
\end{eqnarray}
This system can be solved in terms of elliptic functions. The system has two conservation laws
$$
|\alpha_0|^2 +  2|\beta_0|^2 = E^2,
$$
$$
(\alpha^*_0)^2\beta_0 + (\alpha_0)^2\beta^*_0 = H.
$$
Function $-\displaystyle\frac{iH}{4}$ is a Hamiltonian of  (\ref{WKBleading}). The first conservation law allows us to obtain the following representation for $\alpha_0, \beta_0$ 
$$
\alpha_0 = E\exp\{i\varphi\}\cos(\Psi),\quad \beta_0 = \frac{E}{\sqrt{2}}\exp\{i\psi\}\sin(\Psi).
$$
After this substitution the second conservation law becomes 
\bb
H=\sqrt{2}E^3\cos^2(\Psi)\sin(\Psi)\cos(\Phi),\  \hbox{where}\  \Phi = 2\varphi - \psi. \label{Hamleading}
\ee
System (\ref{WKBleading}) can be written as
\begin{eqnarray}
\varphi' &=& -\frac{E}{2\sqrt{2}}\cos(\Phi)\sin(\Psi),\nonumber \\
\psi' &=& -\frac{E}{2\sqrt{2}}\cos(\Phi)\cos^2(\Psi)\sin^{-1}(\Psi), \label{WKBleading2} \\
\Psi' &=& \frac{E}{2\sqrt{2}}\sin(\Phi)\cos(\Psi),\nonumber
\end{eqnarray}
Using expression (\ref{Hamleading}) allows us to separate system (\ref{WKBleading2}). It yields the separate equation for   $\Psi$
$$
\Psi' = \frac{\sqrt{2E^6\cos^4(\Psi)\sin^2(\Psi)-H^2}}{4E^2\cos(\Psi)\sin(\Psi)}.
$$
The solution of the equation is 
\bb
\int_{u_0}^{\cos(2\Psi)} \frac{du}{\sqrt{G - u^3 -u^2+u}} = -\frac{Et}{2}, \label{solPsi}
\ee 
where $G=\displaystyle\frac{E^6 - 4H^4}{E^6}$. Function $\cos(2\Psi)$ is a bounded  periodic function with respect to $t$. We determine the function $\Phi$  from (\ref{Hamleading}) and integrate the first equation of (\ref{WKBleading2})  
$$
\varphi= \varphi_0 + \frac{H}{4E^2} \int^t \frac{dt}{1+\cos^2(2\Psi)}.
$$
The integrand is a periodic function with respect to $t$ with nonzero average value. It yields the following behaviour 
 $\varphi=O(t),\ t\to \infty.$ The function $\psi$ is determined by $\psi= 2\varphi - \Phi.$ 
Thus we have constructed a family of solutions $\alpha_0, \beta_0$ that depends on four parameters $E, G, u_0, \varphi_0$.
\par
The first correction terms of expansion (\ref{asWKB0}) are determined from the linearized system 
\begin{eqnarray}
i\alpha_1' -\frac{1}{2}(\alpha_1^*\beta_0 + \alpha_0^*\beta_1)&=& -\frac{f}{4}\beta_0\exp\{-it^2\},\nonumber \\
i\beta_1' - \frac{1}{2}\alpha_0\alpha_1 &=& -\frac{f}{4}\alpha_0\exp\{it^2\}. \label{sysWKB2}
\end{eqnarray}
The fundamental matrix $W$ of the homogeneous system that relates to (\ref{sysWKB2}) is formed by the first order derivatives with respect to parameters of solutions for (\ref{WKBleading}). When $t \to \infty$ we obtain the following asymptotic behavior
$$
\pt_{E}\alpha_0 = O(t),\quad \pt_{G}\alpha_0 = O(t),\quad \pt_{u_0}\alpha_0 = O(1),\quad \pt_{\varphi_0}\alpha_0 = O(1). 
$$
The similar formulas are valid for functions $\alpha_0^*, \beta_0, \beta_0^*$. Thus the fundamental matrix contains two columns of the order of $t$ and has the determinant of the order of a constant. 
\par
The solution of the nonhomogeneous system is increasing due to fast oscillations of the right-hand sides. It is convenient to rewrite the right-hand side of the system as a sum of two vectors $g^+\exp\{it^2\}+g^-\exp\{-it^2\}$. 
Here
$$
g^+ = \begin{pmatrix} 0 \\ -\displaystyle\frac{f\beta_0^*}{4} \\ -\displaystyle\frac{f\alpha_0}{4} \\ 0   \end{pmatrix},
\qquad 
g^- = \begin{pmatrix} -\displaystyle\frac{f\beta_0}{4} \\ 0 \\ 0 \\  -\displaystyle\frac{f\alpha_0^*}{4}  \end{pmatrix}.
$$
After integrating twice we obtain
\begin{eqnarray*}
&W&\int_t^\infty W^{-1}g^+\exp\{it^2\}dt =\frac{g^+}{2it}\exp\{it^2\} \\
&+& \frac{W}{4t}(\frac{W^{-1}g^+}{t})'\exp\{it^2\}
-W\int_t^{\infty}\Big[ \Big( \frac{W^{-1}g^+}{2t}\Big)'\frac{1}{2t} \Big]'\exp\{it^2\}dt  \\
&=&\frac{g^+}{it}\exp\{it^2\} + \frac{1}{t}(\frac{g^+}{t})'\exp\{it^2\}+\frac{W}{t}\frac{(W^{-1})'g^+}{t}\exp\{it^2\} \\
&-&W\int_t^{\infty}\Big[ \Big( \frac{W^{-1}g^+}{2t}\Big)'\frac{1}{2t} \Big]'\exp\{it^2\}dt.
\end{eqnarray*}
$Rang W=2$  and any minor of the matrix does not contain terms of the order of $t^2$. Taking into account that  the determinant of matrix $W$ is of order of constant we obtain the order of terms of inverse matrix $W^{-1}$. Their order is $t$. Thus term $\displaystyle\frac{W}{t}\frac{(W^{-1})'g^+}{t}$ has the order of a constant. Another integral can be evaluated by parts and estimated by $O(t^{-1}).$ The similar evaluations are valid for the part of the solution with $g^-.$ It yields that the first correction terms $\alpha_1,\ \beta_1$ are finite.
\par
The next order correction terms are determined from the system of type of (\ref{sysWKB2}). The right-hand sides of the systems are quadratic forms of previous corrections $\alpha_k,\ \beta_k$  and coefficients of (\ref{asZero}). Solutions are bounded. The theorem is proved.

\subsection{A neighborhood of the increasing algebraic solution}

We investigate a neighborhood of the increasing asymptotic algebraic solution $(A_3(t),B_3(t))$ numerically. We solve an initial value problem for big values of $t$. Initial data are close to the algebraic asymptotic solution. Denote by  
$A_3(t;2)$ and $B_3(t;2)$ a sum of two first terms of the asymptotic expansion $(A_3(t),B_3(t))$. We choose $f=12.1$ and 
$$
A|_{t=100}=A_3(100;2)+0.1,\quad B|_{t=100}=B_3(100;2)+0.1;
$$
and solve system (\ref{sys3}) numerically.
\par
The results are presented below. The parameter $t$ goes from 100 to 150. Here we graph on the complex plane $A$ and $B$.
\par
\vskip0.5cm
\begin{tabular}{cc}
\includegraphics[width=5cm,height=4cm]{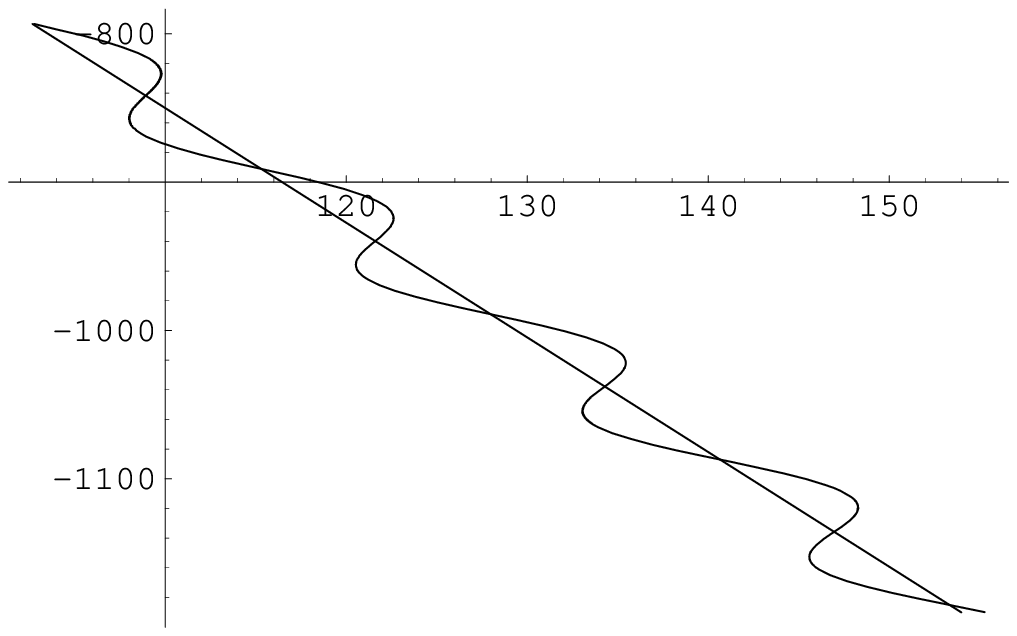}%for DVI-file
& 
\includegraphics[width=5cm,height=4cm]{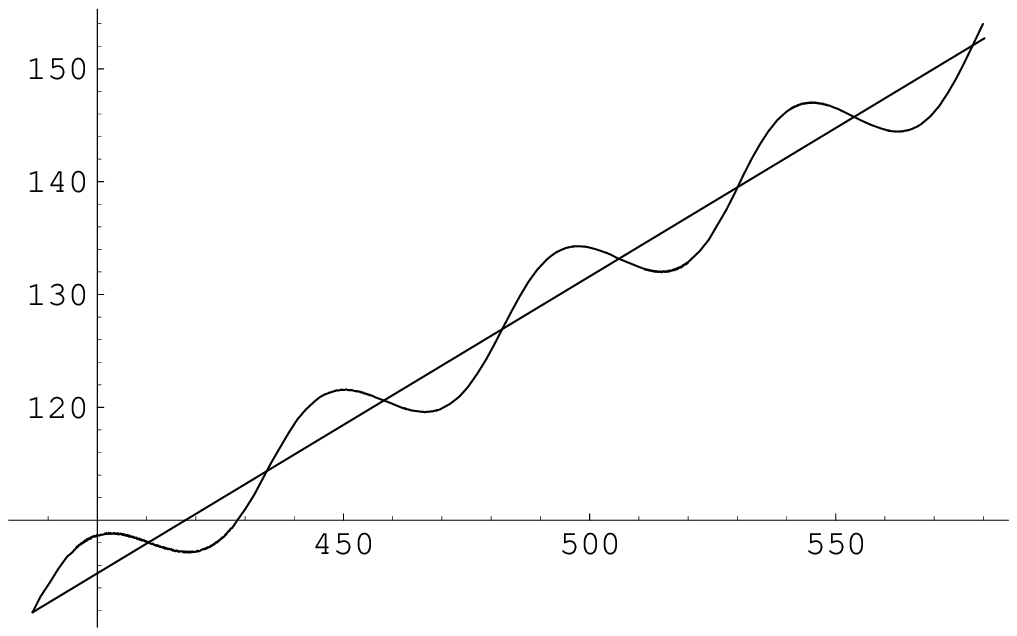} \\ %for DVI-file
\begin{tabular}{c} Fig.3: $A_3(t)$ is a straight line and\\ numerical solution $A(t)$\end{tabular} 
& 
\begin{tabular}{c} Fig.4: $B_3(t)$ is a straight line and \\ numerical solution $B(t)$\end{tabular}
\end{tabular}
\par
\vskip0.5cm
Here we present the graph for comparative difference for numerical solution and algebraic asymptotic solution.
\par
\vskip0.5cm
\begin{tabular}{cc}
\includegraphics[width=5cm,height=4cm]{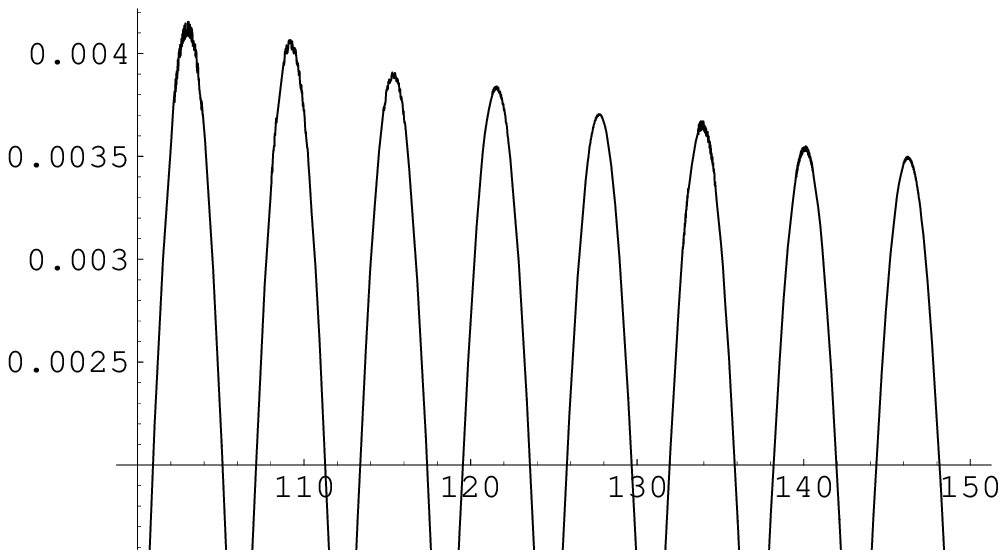}%for DVI-file
& 
\includegraphics[width=5cm,height=4cm]{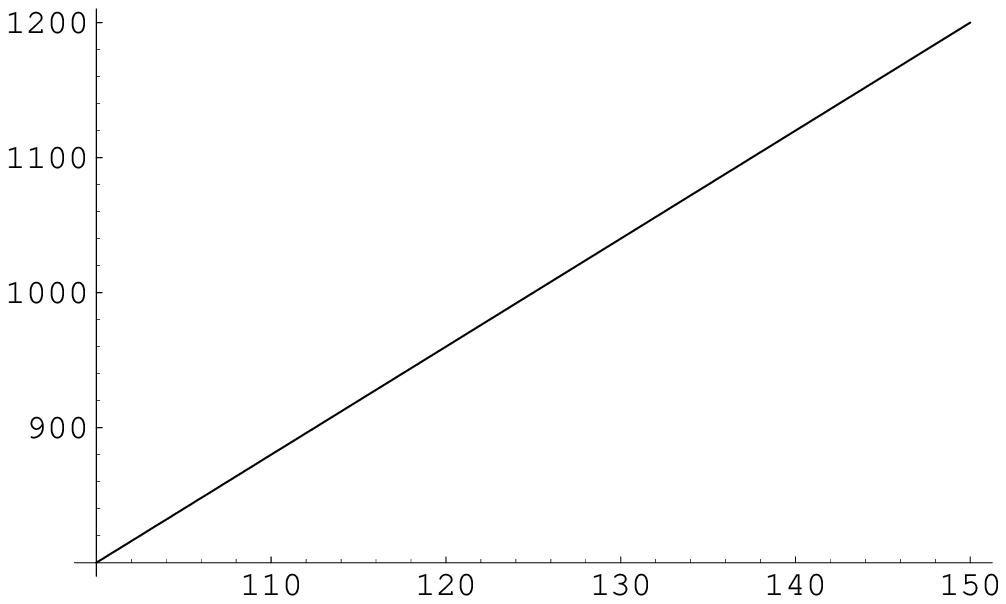} \\ %for DVI-file
\begin{tabular}{c} Fig.5: comparative \\ difference: $\bigg|{A(t)-A_3(t)\over A(t)}\bigg|$\end{tabular}
& 
\begin{tabular}{c} Fig.6: Absolute value of  \\ numerical solution $|A(t)|$.\end{tabular}
\end{tabular}
\par
\vskip0.5cm
The graph of numerical solution  $|B(t)|$ looks the same. The constructed above numerical solution  proves the existence of oscillating solutions near the increasing algebraic asymptotic solution.
\par
We thank Prof. L.A. Kalyakin for stimulating discussions and  Prof. A.D. Bruno for discussions. 
We are also grateful to Anna A. Shchipitsyna, PhD  for helping us to prepare the  presentation of our results.
\par

\end{document}